\documentclass[final,twocolumn,showpacs,showkeys,groupaddress,preprintnumbers,floatfix]{revtex4-1}

\usepackage{dynlearn}

\newcommand{\kB}{\ensuremath{k_\text{B}}\xspace}

\newcommand{\pup}{\p{\uparrow}}
\newcommand{\pdown}{\p{\downarrow}}

\newcommand{\spin}[1]{\ensuremath{\sigma_{#1}}\xspace}
\newcommand{\spinrv}[1]{\ensuremath{\boldsymbol{\sigma}_{#1}}\xspace}
\newcommand{\spins}{\ensuremath{\sigma}\xspace}
\newcommand{\spinsrv}{\ensuremath{\boldsymbol{\sigma}}\xspace}
\newcommand{\spinsminus}[1]{\ensuremath{\spins_{\setminus #1}}\xspace}
\newcommand{\spinsminusrv}[1]{\ensuremath{\spinsrv_{\setminus #1}}\xspace}

\newcommand{\lattice}{\ensuremath{\mathcal{L}}\xspace}
\newcommand{\latticesize}{\ensuremath{\left| \lattice \right|}\xspace}

\newcommand{\PATTERN} {bound\xspace}

\newcommand{\Tc}{\ensuremath{T_{c}}\xspace}
\newcommand{\Tb}{\ensuremath{T_{b}}\xspace}
\newcommand{\Tgl}{\ensuremath{T_{\mathcal{T}_{gl}}}\xspace}

\colorlet{H1_color} {black}
\colorlet{hmu_color}{blue}
\colorlet{rho_color}{red}
\colorlet{rmu_color}{orange}
\colorlet{bmu_color}{green!66!black}
\colorlet{qmu_color}{purple}

\begin{document}

\def\ourTitle{%
  Anatomy of a Spin:\\
  The Information-Theoretic Structure of Classical Spin Systems
}

\def\ourAbstract{%
  Collective organization in matter plays a significant role in its expressed physical properties. Typically, it is detected via an order parameter, appropriately defined for each given system's observed emergent patterns. Recent developments in information theory, however, suggest quantifying collective organization in a system- and phenomenon-agnostic way: decompose the system's thermodynamic entropy density into a \emph{localized entropy}, that solely contained in the dynamics at a single location, and a \emph{\PATTERN entropy}, that stored in space as domains, clusters, excitations, or other emergent structures. We compute this decomposition and related quantities explicitly for the nearest-neighbor Ising model on the 1D chain, the Bethe lattice with coordination number $k=3$, and the 2D square lattice, illustrating its generality and the functional insights it gives near and away from phase transitions. In particular, we consider the roles that different spin motifs play (in cluster bulk, cluster edges, and the like) and how these affect the dependencies between spins.
}

\def\ourKeywords{%
  Ising spin model, thermodynamic entropy density, dual total correlation,
  entropy rate, elusive information, enigmatic information, predictable
  information rate, complex system
}

\hypersetup{
  pdfauthor={Ryan G. James},
  pdftitle={\ourTitle},
  pdfsubject={\ourAbstract},
  pdfkeywords={\ourKeywords},
  pdfproducer={},
  pdfcreator={},
}

\author{Vikram S. Vijayaraghavan}
\email{vsvijayaraghavan@ucdavis.edu}
\affiliation{Complexity Sciences Center and Physics Department,
University of California at Davis, One Shields Avenue, Davis, CA 95616}

\author{Ryan G. James}
\email{rgjames@ucdavis.edu}
\affiliation{Complexity Sciences Center and Physics Department,
University of California at Davis, One Shields Avenue, Davis, CA 95616}

\author{James P. Crutchfield}
\email{chaos@ucdavis.edu}
\affiliation{Complexity Sciences Center and Physics Department,
University of California at Davis, One Shields Avenue, Davis, CA 95616}

\date{\today}
\bibliographystyle{unsrt}

\title{\ourTitle}

\begin{abstract}
\ourAbstract
\end{abstract}



\preprint{\sfiwp{15-10-042}}
\preprint{\arxiv{1510.08954 [cond-mat.stat-mech]}}

\title{\ourTitle}
\date{\today}
\maketitle

%


\setstretch{1.1}

\section{Introduction}
\label{sec:introduction}

Collective behavior underlies a vast array of fascinating phenomena, many of critical importance to contemporary science and technology. Unusual material properties---such as superconductivity, metal-insulator transitions, and heavy fermions---have been attributed to collective behavior arising from the compounded interaction of system components~\cite{antonov2011electronic}. Collective behavior is by no means limited to complex materials, however: the behavior of financial markets~\cite{mantegna1999introduction}, epileptic seizures~\cite{raiesdana2008complexity} and consciousness~\cite{maki2013disruption}, and animal flocking behavior~\cite{couzin2007collective} are all now also seen as examples. And, we now appreciate that it appears most prominently via phase transitions~\cite[and references therein]{barnett2013information}.

Operationally, collective behavior is detected and quantified using correlations or, more recently, by mutual information \cite{Cove06a}, estimated either locally via pairwise component interactions or globally. Exploring spin systems familiar from statistical mechanics, here we argue that these diagnostics can be substantially refined to become more incisive tools for quantifying collective behavior. This is part of the larger endeavor of discovering ways to automatically detect collective behavior and the emergence of organization \cite{Crut12a}.

Along these lines, much effort has been invested to explore information-theoretic properties of the Ising model of statistical mechanics. Both Shaw~\cite{Shaw84} and Arnold~\cite{Arno96} studied information in 1D spin-strips within the 2D Ising model. Feldman and Crutchfield~\cite{crutchfield1997statistical,feldman2003structural,Feld08a} explored several generalizations of the excess entropy, a well known mutual information measure for time series, in 1D and 2D Ising models as a function of coupling strength, showing that they were sensitive to spin patterns and can be used as a generalized order parameter. In one of the more thorough-going studies to date, Lau and Grassberger~\cite{lau2013information} computed the excess entropy using adjacent rings of spins to probe criticality in the 2D cylindrical Ising model. Barnett \etal~\cite{barnett2013information} tracked information flows in a kinetic Ising system using both the pairwise and global mutual informations (generalized as the total correlation \cite{Jame11a}, it appears below) and the transfer entropy \cite{Schr00a}. Abdallah and Plumbley~\cite{abdallah2012measure} employed an extensive version of the dual total correlation \cite{Jame11a} on small, random spin glasses. Despite their successful application, the measures used to monitor collective phenomena in these studies were motivated information theoretically rather than physically, and so their thermodynamic relevance and structural interpretation have remained unclear.

To address this we decompose the thermodynamic entropy density for spin systems into a set of information measures, some already familiar in information theory~\cite{verdu2006erasure}, complex systems~\cite{james2014chaos}, and elsewhere~\cite{abdallah2012measure}. For example, one measure that falls out naturally---the \PATTERN entropy---is that part of the thermodynamic entropy accounting for entropy shared \emph{between} spins. This is in contrast to monitoring collective behavior as the difference between the thermodynamic entropy and the entropy of a hypothetical distribution over uncorrelated spins---the total correlation mentioned above. In this way, the decomposition provides a physical basis for informational measures of collective behavior, in particular showing how the measures lead to interpretations of emergent structure in system configurations.

To make this argument and illustrate its consequences, Section\nobreakspace \ref {sec:spin_entropy} first defines notation and lays the groundwork for our decomposition. Section\nobreakspace \ref {sec:decomposition} then derives the decomposition. Though this decomposition is given for regular spin lattices, in principle it is meaningful for any system with a well defined thermodynamic entropy. Section\nobreakspace \ref {sec:results} outlines the computational methods for estimating the resulting quantities, using them to interpret emergent organization in the nearest-neighbor ferromagnetic Ising model in one dimension (Section\nobreakspace \ref {subsec:1d_ising}), in the Bethe lattice with coordination number $k=3$ (Section\nobreakspace \ref {subsec:bethe_ising}), and in the two-dimensional square Ising lattice (Section\nobreakspace \ref {subsec:2d_ising}). Having established the different types of entropy in spin systems and connections to their underlying physical structures, we conclude by suggesting applications and future directions.

\section{Spin Entropies}
\label{sec:spin_entropy}

We write \spins to denote a configuration of spins on a lattice \lattice, \spin{i} for the particular spin state at lattice site $i \in \lattice$, and \spinsminus{i} the collection of all spin states in a configuration excluding \spin{i} at site $i$. The random variable over all possible spin configurations is denoted \spinsrv, for a particular spin variable \spinrv{i}, and for all spin variables but the one at site $i$, \spinsminusrv{i}. As a shorthand, we write $\spins \in \spinsrv$ and similar to mean indexing into the event space of the random variable \spinsrv.

We study the ferromagnetic spin-$1/2$ Ising model with nearest-neighbor interactions in the thermodynamic limit whose Hamiltonian is given by:
\begin{align} \label{eq:hamiltonian}
  \mathcal{H}(\spins) = -J \sum_{\langle i, j \rangle} \spin{i} \spin{j}
  ~,
\end{align}
where $\langle i, j \rangle$ denotes all pairs $(i, j)$ such that the sites $i$ and $j$ are directly connected in \lattice and the interaction strength $J$ is positive. We assume the system is in equilibrium and isolated. And so, the probability of configuration \spins occurring is given by the Boltzmann distribution:
\begin{align}
  \p{\spins} = \frac{1}{Z}e^{-\mathcal{H}(\spins) / \kB T}~,
\end{align}
where $Z$ is the partition function.

The \emph{Boltzmann entropy} of a statistical mechanical system assumes the constituent degrees of freedom (here spins) are uncorrelated. To determine it, we use the \emph{isolated spin entropy}:
\begin{align*}
  \H{\spinrv{0}} = - \pup \log_2{\pup} - \pdown \log_2{\pdown}
  ~,
\end{align*}
where $\pup = (1+m)/2$ is the probability of a spin being up, $\pdown = (1-m)/2$ is the probability of a spin being down, and $m = (\#\uparrow - \#\downarrow)/\latticesize$ is average magnetization in a configuration. The site index $0$ was chosen arbitrarily and represents any single spin in the lattice. The system's Boltzmann entropy is then the extensive quantity $H_\text{B} = \latticesize \cdot \H{\spinrv{0}}$.

As Jaynes~\cite{jaynes1965gibbs} emphasizes, though correct for an ideal gas, $H_\text{B}$ is not the empirically correct system entropy if a system develops internal correlations. In our case, if spins are correlated across the lattice, one must consider entire configurations, leading to the \emph{Gibbs entropy}:
\begin{align} \label{eq:HGibbs}
  \H{\spinsrv} = - \sum_{\spins \in \spinsrv} \p{\spins} \log_2 \p{\spins}
\end{align}
and the \emph{thermodynamic entropy density} $h$, the entropy per spin when the entire lattice is considered:
\begin{align}
  h = \frac{\H{\spinsrv}}{\latticesize}
  ~.
  \label{eq:entropy_density}
\end{align}
Note that we dropped the factor \kB and used the base-$2$ logarithm. Our use of the letter $h$ as opposed to $s$ is meant to reflect this multiplicative constant difference. It is easy to see that $\H{\spinsrv} \leq H_\text{B}$. Thus, the Boltzmann entropy is typically an overestimate of the true thermodynamic entropy.

More to the point, in the thermodynamic limit the Boltzmann entropy $H_\text{B}$ and Gibbs entropy $\H{\spinsrv}$ differ substantially. As Jaynes shows~\cite{jaynes1965gibbs}, the difference directly measures the effect of internal correlations on total energy and other thermodynamic state variables. Moreover, the difference does not vanish, rather it increases proportionally to system size \latticesize.

Before leaving the differences between entropy definitions, it is important to note that Boltzmann's more familiar definition---$S = \kB \ln W$---via the number $W$ of microstates associated with a given thermodynamic macrostate is consistent with Gibbs' definition \cite{jaynes1965gibbs}. This follows from Shannon's deep insight on the \emph{asymptotic equipartition property} that $2^{h \latticesize}$ measures the volume of \emph{typical} microstates: the set of almost-equiprobable configurations that are simultaneously most numerous \emph{and} capture the bulk of the probability---those that are typically realized. (See Refs. \cite[Sec. 21]{Shan48a}, \cite[Ch.~3]{Cove06a} and \cite{grandy2008entropy,ruelle1969rigorous}.) Thus, Boltzmann's $W$ should be interpreted as the size (phase space volume) of the typical set associated with a particular thermodynamic macrostate. Given this agreement, we focus Gibbs' approach. Though, as we now note, the contrast between Gibbs' entropy $\H{\spinsrv}$ and Boltzmann's $H_\text{B}$ leads directly to our larger goals.

\section{Decomposing a Spin's Thermodynamic Entropy}
\label{sec:decomposition}

Since we are particularly interested here in monitoring the appearance of internal correlations, the difference between the Boltzmann and Gibbs entropies suggests itself as our first measure of a system's internal organization. If each spin were independent, $\H{\spinrv{0}} = h$. For example, this is true at $T=\infty$ for the Ising model and for site percolation models \cite{Grim99a}. If  there are correlations between spins, however, then $h < \H{\spinrv{0}}$, as just noted for the extensive quantities. Their difference is the \emph{total correlation density}~\cite{Jame11a}:
\begin{align}
\label{eq:TotalCorr}
  \rho & = \H{\spinrv{0}} - h \\
       & = \frac{\T{\spinsrv}}{\latticesize}
  \nonumber
  ~,
\end{align}
\T{\spinsrv} here is the Kullback-Leibler divergence~\cite{Cove06a} between the distribution over entire configurations and the product of isolated-spin marginal distributions. Now called the \emph{total correlation}~\cite{Jame11a}, this measure of internal correlation was wholly anticipated by Gibbs, as Jaynes notes \cite[Eq.~(4)]{jaynes1965gibbs}. Since $\rho$ vanishes only when each spin is independent, one might consider it a measure of pattern or structure in a spin system. While this is certainly reasonable as an operational definition, our next step decomposes both $h$ and $\rho$ into more nuanced quantities.

Continuing, we recast the Gibbs thermodynamic entropy \H{\spinsrv} as the sum of two nonnegative terms; starting from the configuration entropy in its standard statistical form Eq.\nobreakspace \textup {(\ref {eq:HGibbs})} and manipulating it into two terms:
\begin{align}
  \H{\spinsrv} &= - \sum_{\spins \in \spinsrv} \p{\spins} \log_2 \left[ \p{\spins} \frac{\displaystyle\prod_{i=1}^{\latticesize} \p{\spin{i} \mid \spinsminus{i}}} {\displaystyle\prod_{i=1}^{\latticesize} \p{\spin{i} \mid \spinsminus{i}}} \right] \nonumber \\
  &= \R{\spinsrv} + \B{\spinsrv}
  ~,
\label{eq:ConfigEntropyDecomp}
\end{align}
where \R{\spinsrv} and \B{\spinsrv}, known as the \emph{residual entropy} and \emph{dual total correlation}~\cite{verdu2006erasure,abdallah2012measure,Jame11a}, are the following measures:
\begin{align}
  \R{\spinsrv}
        &= \sum_{i=1}^{\latticesize} \left[ -\sum_{\spins \in \spinsrv}
		\p{\spins} \log_2 \p{\spin{i} \mid \spinsminus{i}} \right] \nonumber \\
        &= \sum_{i=1}^{\latticesize} \H{\spinrv{i} \mid \spinsminusrv{i}},
  \label{eq:rmu}
\end{align}
and
\begin{align}
  \B{\spinsrv} &= -\sum_{\spins \in \spinsrv} \p{\spins} \log_2 \frac{\p{\spins}}{\displaystyle\prod_{i=0}^{\latticesize} \p{\spin{i} \mid \spinsminus{i}}} \\
  &= \H{\spinsrv} - \sum_{i=0}^{\latticesize} \H{\spinrv{i} \mid \spinsminusrv{i}} \nonumber
	~.
\end{align}
Note that both \R{\spinsrv} and \B{\spinsrv} are nonnegative and bound from above by \H{\spinsrv}. Spatial densities are denoted in lower case: $h = \nicefrac{\H{\spinsrv}}{\latticesize}$, $r = \nicefrac{\R{\spinsrv}}{\latticesize}$, and $b = \nicefrac{\B{\spinsrv}}{\latticesize}$. We consider their thermodynamic limit, where $\latticesize \to \infty$.

Similarly, we can begin with the total correlation of spins \T{\spinsrv} and break it in to two terms:
\begin{align}
  \T{\spinsrv} &= \sum_{\spins \in \spinsrv} \p{\spins} \log_2 \frac{\p{\spins}}{\displaystyle \prod_{i=1}^{\latticesize} \p{\spin{i}}} \\
  &= \B{\spinsrv} + \Q{\spinsrv}
  ~,
\end{align}
where \B{\spinsrv} is as above. Though perhaps not clear here, \B{\spinsrv} is naturally a component of \T{\spinsrv}; see Ref.~\cite[Figs.~7c~and~7d]{Jame11a}. \Q{\spinsrv} is the \emph{enigmatic information}~\cite{Jame11a}:
\begin{align}
  \Q{\spinsrv} &= \sum_{\spins \in \spinsrv} \p{\spins} \log_2 \frac{\p{\spins}^2}{\displaystyle \prod_{i=1}^{\latticesize} \p{\spin{i}} \prod_{j=1}^{\latticesize} \p{\spin{j} \mid \spinsminus{j}}} \\
  &= \T{\spinsrv} - \B{\spinsrv} \nonumber
  ~,
\end{align}
and again, we consider the spatial density $q = \nicefrac{\Q{\spinsrv}}{\latticesize}$ in the thermodynamic limit.

\begin{figure}
  \contourlength{0.1em}
  \includegraphics{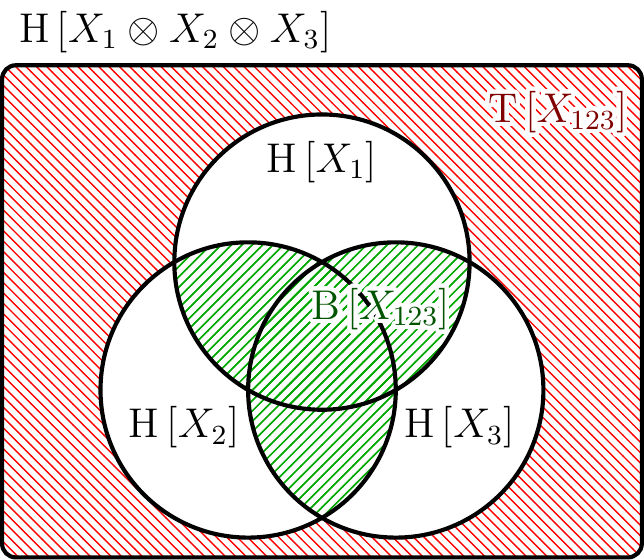}
  \caption{
    Total correlation \T{X} (north-west hashing) and dual total correlation \B{X} (north-east hashing) for three spins $X_1$, $X_2$, and $X_3$ depicted via a $3$-variable Venn diagram~\cite{reza1961introduction} in which areas correspond to the magnitudes of information measures. The joint entropy $\H{X_{123}}$ is the union of the three single-spin marginal entropies $\H{X_i}$ (full circles). The outer rectangle represents an entropy $\H{X_1 \otimes X_2 \otimes X_3}$ with all dependencies between spins removed; each marginal entropy $\H{X_i}$ still matches its original marginal entropy, but does not overlap that of the other spins. This corresponds to the distribution whose Gibbs entropy is the Boltzmann entropy of $X_{123}$. The diagram directly demonstrates that \T{X} measures dependence as a distance from the external reference $\H{X_1 \otimes X_2 \otimes X_3}$. Whereas, \B{X} measures with respect to an internal reference $\H{X_{123}}$. The unshaded region is residual entropy \R{X_{123}}.
  }
  \label{fig:dependencies}
\end{figure}

It can be difficult to initially intuit the difference between \T{\spinsrv} and \B{\spinsrv}. Both measure dependencies among spins, but against different reference ensembles. \T{\spinsrv} is the difference between the Boltzmann and Gibbs entropies (Eq. (\ref{eq:TotalCorr})) and, therefore, quantifies the spin distribution's deviation from a hypothetical distribution in which each spin is independent from the others. Since \R{\spinsrv} is the amount of actual independent entropy in the spin distribution, \B{\spinsrv} quantifies the amount of spin-spin dependency in configurations not in reference to a hypothetical independent-spin distribution, but rather to the actually realized amount of independence. This difference is pictorially represented in Fig.\nobreakspace \ref {fig:dependencies}.

Let us now interpret the physics captured by these entropy components. First, \H{\spinrv{0}} quantifies the entropy per spin ignoring any dependencies (\eg correlations) between spins. The thermodynamic entropy density $h$, however, is the entropy per spin including dependencies. Continuing in this way, from Eq.\nobreakspace \textup {(\ref {eq:rmu})} $r$ is the entropy per spin remaining after these dependencies have been factored out, meaning it is the average amount of \emph{independent} entropy---the entropy per spin when we know the state of the other spins in the lattice. With this logic, $b$ is that portion of the thermodynamic entropy density coming from dependencies and is, therefore, the average amount of \emph{dependent} entropy per spin. As we shall see, in comparison to $\rho$, $b$ provides a complementary, quantitatively different, and more physically grounded approach to quantifying dependencies among spins.

\begin{figure}
  \centering
  \includegraphics{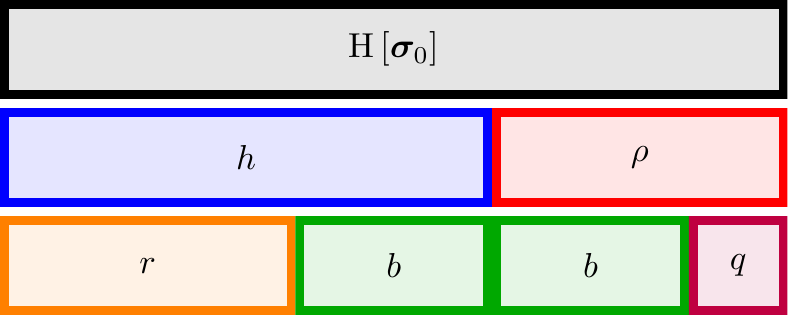}
  \caption{%
    Decomposition of the isolated-spin (Boltzmann) entropy \H{\spinrv{0}}. First, into the Gibbs thermodynamic entropy density $h$ and the total correlation $\rho$. Second, the thermodynamic entropy into the localized density $r$ and the dual total correlation information density $b$, and the total correlation density $\rho$ into $b$ and the enigmatic information density $q$.
  }
  \label{fig:decomposition}
\end{figure}

This completes, in effect, our decomposition of the isolated-spin entropy \H{\spinrv{0}}, shown schematically in Fig.\nobreakspace \ref {fig:decomposition}. To take stock, let's back up a bit. Recall that $\rho$ is the difference between \H{\spinrv{0}}, which ignores the dependencies between the spins, and the thermodynamic entropy density $h$. That is, $\rho$ is the difference between the isolated-spin entropy and how much entropy ($h$) there actually is. Whereas, $b$ is the difference between the thermodynamic entropy $h$ and the amount of independent entropy $r$ there is. In addition, $b$ is also part of $\rho$~\cite{Jame11a}, leaving $q$ which therefore quantifies potential dependencies (explained shortly) not present in the thermodynamic entropy $h$.

Let's explore the consequences of this decomposition. Although $r$ and $b$ are components of the thermodynamic entropy density $h$---itself a measure of the irreducible per-spin randomness---individually they have very different meanings that give insight into the interplay of spin thermalization and spin ordering. The first, $r$, is the average amount of unshared entropy, quantifying randomness remaining in a single spin given the exact configuration of the rest of the lattice. For this reason we refer to it as the \emph{localized entropy}. In the limit of high and low temperature, $r$'s behavior is intuitive. At low temperature, spins in the lattice are entirely aligned, and so there is no randomness in individual sites given the global alignment. At high temperatures, each spin is independent, and so there is full randomness in a site, even knowing its surroundings.

The other quantity, $b$, is the entropy remaining in a spin when the localized entropy is deducted, and so it captures the thermodynamic entropy shared between a spin and the rest of the lattice. We refer to it as the \emph{\PATTERN entropy}. Its limits are similarly intuitive. At low temperatures there is no randomness to share, and so $b = 0$. Similarly, at high temperatures all randomness is localized, and so no entropy can be shared and \PATTERN entropy also vanishes. At intermediate temperatures, though spins are in flux, a spin's context does at least in part influence its alignment and so $b > 0$. We refer to the temperature at which $b$ is maximized by \Tb.

Finally, we consider the enigmatic information density $q$. As a component of $\rho$ it also reflects dependencies, but in a way different from $b$. The \PATTERN entropy captures dependencies that are also part of the thermodynamic entropy, but $q$ captures those that are not. As we shall see in Section\nobreakspace \ref {sec:discussion}, this means it is sensitive to the interior of clusters. At low temperatures, \ie, temperatures below \Tc, we expect $q$ to be small, since \H{\spinrv{0}} is small. At high temperatures, there are no dependencies between spins, and so we also expect $q$ to be small. At intermediate temperatures, just like $b$, we expect $q$ to take larger values.

The nonzero values of $\rho$, $b$, and $q$ are all driven by Ising spin clustering. Consider for the moment randomly permuting the sites in the lattice, resulting in a site percolation model with site occupancy probability given by a simple function of the magnetization of the original Ising lattice. Then, the spin orientation at any particular site is now statistically independent of its neighbors, and so $\rho = b = q = 0$, and therefore $\H{\spinrv{0}} = h = r$. There are, however, clusters in this permuted lattice---in fact, there may exist a giant cluster. These clusters are driven purely by the lattice topology. We therefore conclude that $\rho$, $b$, and $q$ are each sensitive to differing aspects of how the pairwise Ising Hamiltonian affects the shape and distribution of spin clusters.

To summarize, the disorder generated at each site is measured by the Gibbs thermodynamic entropy density $h$. And, one portion ($b$) participates in spatial organization and the other portion ($r$) does not. Thus, $b$ is key to monitoring the emergence of spatial structures during pattern formation dynamics.

\section{Results}
\label{sec:results}

To explore what the entropy components reveal, we calculate them for the ferromagnetic spin-\half Ising model on one- (1D) and two-dimensional (2D) square lattices and on the Bethe lattice. These results demonstrate that our intuitions regarding the behaviors of the information measures are correct, but also raise subtle questions regarding the details of exactly what they capture.

As part of this, we report analytical results for the 1D chain and the Bethe lattice and show that they match results from simulation. And, we use analytic results whenever possible for the 2D lattice. For example, we report analytical results for $h$, $\H{\spinrv{0}}$, and $\rho$.  Then, only $r$ needs to be estimated from simulation and that result is then combined with analytic quantities to compute $b$ and $q$. To provide a comparison, estimating $b$ purely from simulation requires accurate $h$ and $r$. In this, we make use of the global Markovian property of spin lattices~\cite{goldstein1990entropy}. To obtain $r$ we need only condition a spin on all its nearest neighbors---spins directly coupled via the Hamiltonian of Eq.\nobreakspace \textup {(\ref {eq:hamiltonian})}. For $h$ estimated purely from simulation, we use Ref. \cite{schlijper1989two}'s method. In both cases, Ref.~\cite{schurmann2015note}'s entropy estimator $\widehat{H}_2$ was used to decrease the number of configuration samples while still providing convergence to known analytic results. Overall, when comparing and when analytic results are available, simulation results match to within standard error bars smaller than plot line widths. All simulation results were obtained using the Wolff algorithm~\cite{wolff1989collective}. In summary, in the reported results for those several quantities that required estimated or partial estimation from simulations, sufficient care was taken that the statistical errors fall far below the level needed to support the main conclusions.

\subsection{1D Ising Model}
\label{subsec:1d_ising}

Simple Ising models have been broadly adapted to understand simple collective phenomena in fields ranging from surface physics \cite{maniwa2003,zimmermann2000} and biophysics \cite{bruno1960,wartell1985} to economics \cite{Durl99a}. The Ising model on a one-dimensional lattice ($\lattice = \mathbb{Z}$) can be fully analyzed in closed form. Unfortunately, its emergent patterns are rather constrained; for example, it does not exhibit a phase transition~\cite{pathria1996}. However, the exact solutions provide an important benchmark and so it is a necessary candidate with which to start and provides a base from which to generalize~\cite{baxter1982,pfeuty1979,Feld98a,yilmaz2005}. Analytic results were computed using a transfer matrix approach combined with the aforementioned Markovian property. For comparison, simulations were performed on a lattice of size $N = 1024$.

\begin{figure}
  \includegraphics{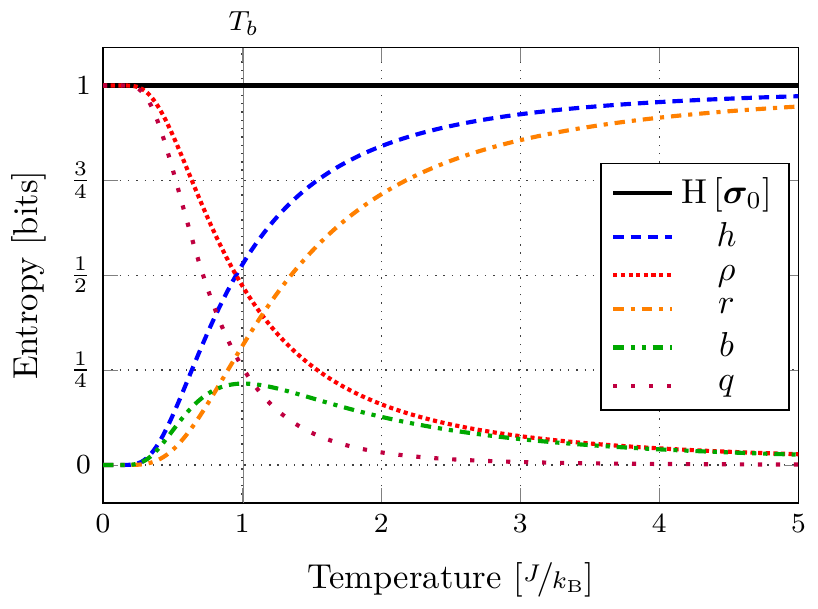}
\caption{Boltzmann and Gibbs thermodynamic entropy decompositions for the 1D,
	nearest neighbor, ferromagnetic spin-$1/2$ Ising model: Gibbs entropy
	density $h$ and the localized entropy density $r$ both monotonically
	increase with temperature, but the \PATTERN entropy density $b$ is maximal
	near $\Tb \approx 1.0117~J/\kB$. Below this temperature, the dominant
	contribution to the Gibbs thermodynamic entropy switches to the entropy
	shared between nearby spins.
  }
  \label{fig:ising_1d}
\end{figure}

Independent of sporting a phase transition or not, the argument that $b$ maximizes between temperature extremes still holds and Fig.\nobreakspace \ref {fig:ising_1d} verifies this. The \PATTERN entropy $b$ reaches a maximal value at $T_b \approx 1.0117~J/\kB$ and this is not (and cannot be) associated with a phase transition. At lower temperatures, spin-spin shared entropy $b$ is the dominant contributor to the Gibbs thermodynamic entropy density $h = r + b$. Whereas at higher temperatures, the localized entropy $r$ is the dominant contributor. Similarly, at low temperatures $q$ is the dominant contributor to $\rho$, and $b$ is its dominant contributor at high temperatures. Section\nobreakspace \ref {sec:discussion} below isolates why $b$ peaks where it does and to which spin-configuration features each measure is sensitive. For now, let's continue with system informational phenomenology.

\subsection{Bethe Lattice Ising Model}
\label{subsec:bethe_ising}

What role does the underlying lattice topology play in determining the balance between local randomness and spatially shared information that equilibrium configurations achieve? Spins on the Bethe lattice, in which \lattice is an infinite Cayley tree with coordination number $k$, is an ideal candidate since the absence of loops makes it possible to compute quantities analytically. Moreover, spin configurations exhibit a phase transition at critical temperature \cite{baxter1982}:
\begin{align}
  \Tc = 2 J \left[ \kB \ln \left( \frac{k}{k-2} \right) \right]^{-1}
  ~.
\end{align}
We analytically calculated the entropy decomposition for a Bethe lattice with coordination number $k$, the details of which appear in Appendix\nobreakspace \ref {sec:bethe_analytic}. Simulation results, matching the analytic to high precision, were performed on a $1,000$ node random $3$-regular graph~\cite{bollobas2001}, which has no boundary and is locally Bethe lattice-like.

\begin{figure}
  \includegraphics{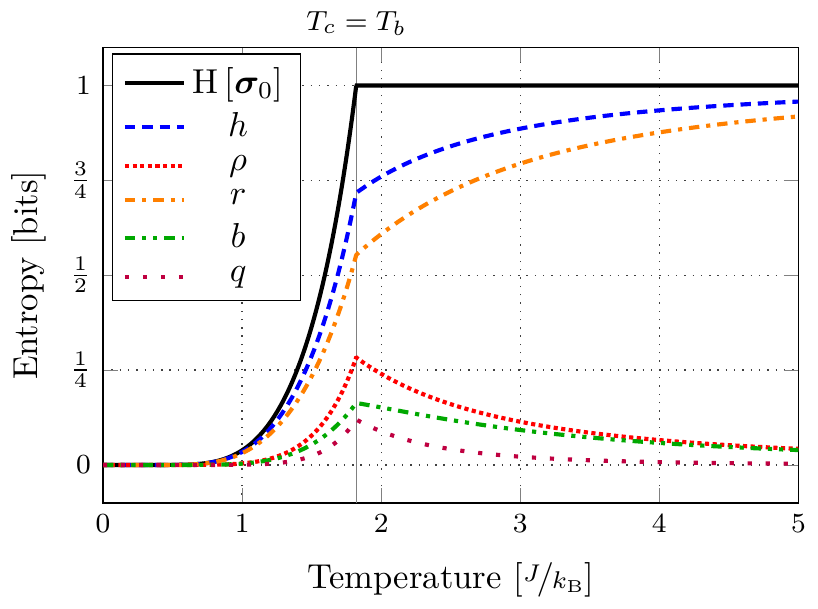}
  \caption{%
    Thermodynamic entropy decompositions for the Ising model on a Bethe lattice. By far and away, and unlike the 1D spin lattice, the individual-spin disorder $r$ is the dominant contributor to Gibbs entropy density $h$ over the entire temperature range. Of the information generated, relatively little ($b$) is stored in spatial patterns.
}
  \label{fig:ising_bethe}
\end{figure}

Figure\nobreakspace \ref {fig:ising_bethe} presents the results for a Bethe lattice with coordination number $k=3$, though other $k$s behave similarly. Interestingly, all information measures have a discontinuity in their first derivatives, and this happens at the phase transition at $\Tc$. Furthermore, the \PATTERN entropy $b$ is maximized there: $\Tb = \Tc \approx 1.8205~J/\kB$. Opposite the 1D lattice, at small $T$ the dominant contributor to Gibbs entropy $h$ is the local randomness $r$. Thus, not only does the change in lattice topology induce a phase transition, but it also inverts the informational components' contributions to thermodynamic entropy density. This, in turn, indicates a rather different underlying mechanism that supports entropy production in the low temperature regime. In the 1D case entropy is spatially extended and configurations of spins are dominated by large clusters. Whereas, in the Bethe lattice deviations from uniformity come in the form of isolated spins differing from their surroundings. Also, unlike 1D, the spin-spin shared entropy $b$ is the dominant contributor to $\rho$ at both low and high temperatures.

\subsection{2D Ising Model}
\label{subsec:2d_ising}

\begin{figure}
  \includegraphics{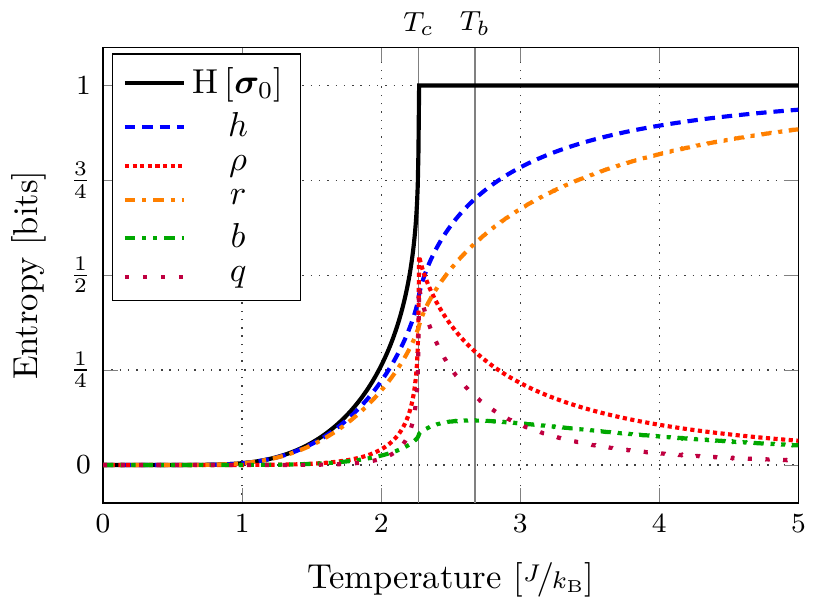}
  \caption{%
    Thermodynamic entropy decompositions for the 2D, nearest neighbor, ferromagnetic Ising model. Curves are from numerical simulation with sufficient size that standard errors are much smaller than the line widths.  As in 1D, the entropy density $h$ and the localized entropy density $r$ monotonically increase with temperature. Here, also, the \PATTERN entropy density $b$ reaches a maximal value at a nonextremal temperature: near $\Tb \approx 2.675~J/\kB$, but this peak value does not occur at the critical temperature $\Tc \approx 2.2692~J/\kB$, where domain sizes become scale-free.
  }
  \label{fig:ising_2d}
\end{figure}

Unlike its 1D sibling, the nearest-neighbor ferromagnetic Ising model in two dimensions ($\lattice = \mathbb{Z}^2$) has a phase transition at a finite critical temperature $\Tc = \nicefrac{2}{\ln\left(1 + \sqrt{2}\right)} \approx 2.2692~J/\kB$. In the 2D case, although Onsager's solution lets us calculate \H{\spinrv{0}}, $h$, and $\rho$~\cite{baxter1982}, we do not have an analytic form for $r$ and so the curves for $r$, $b$, and $q$ in Fig.\nobreakspace \ref {fig:ising_2d} partly rely on estimates from simulation, as explained above. The simulations were conducted on a $128 \times 128$ lattice and quantities averaged over $200,000$ configuration updates. Again, the resulting standard error bars were smaller than the plotted line widths.

Phase transition aside, the behaviors of $h$, $r$, and $b$, seen in Fig.\nobreakspace \ref {fig:ising_2d}, are qualitatively similar to those in the 1D lattice. However, unlike 1D, at low temperatures $r$ is the dominant contributor to $h$, similar to the Bethe lattice below its transition. Also, paralleling the Bethe lattice, $b$ is the dominant contributor to $\rho$ at both low and high temperatures. Like the 1D system, \PATTERN entropy is maximized within the disordered phase at $\Tb \approx 2.675~J/\kB$, \emph{above} $\Tc$. Let us now turn to what configurational structures lead to the overall behaviors of these measures and how they contribute to the Gibbs entropy density $h$.

\section{Discussion}
\label{sec:discussion}

At first, it is striking that \Tb is not generically identical to \Tc. To understand why, we need to investigate to which spin-configuration motifs the various measures are sensitive. To accomplish this, we take a local approach. Each of the measures examined above is an average density, which is important when discussing the lattice as a whole and its bulk thermodynamic properties. However, each spin configuration motif contributes differently to this average. For example, an up spin surrounded by down spins contributes to the measures differently than an up spin surrounded by other up spins.

\subsection{Motif Entropies}
\label{sec:motif}

To quantify this, we appeal to spatial- and temporal-local forms of the averaged densities considered so far. This is possible on square lattices $\lattice = \mathbb{Z}^d$, due to the existence of conditional forms of the entropy density; see Ref.~\cite[Thm.~2.9]{marcus2013computing}. The conditional forms for each of the component quantities are configuration-weighted averages over quantities of the form $\log_2{\left(\bullet\right)}$~\cite{Jame11a}. For example, the thermodynamic entropy density of Eq.\nobreakspace \textup {(\ref {eq:entropy_density})} can be shown to be~\cite{marcus2013computing}:
\begin{align}
  h = \H{\spinrv{i} \mid \overleftarrow{\spinrv{i}}}
    &= -\sum_{\spins \in \spinsrv} \p{\spins} \log_2 \p{\spin{i} |
	\overleftarrow{\spin{i}}} \nonumber \\
    &= \phantom{-}\sum_{\spins \in \spinsrv} \p{\spins} h_i(\spins)
\label{eq:localentropy}
\end{align}
where $h_i(\spins) = - \log_2 \p{\spin{i} | \overleftarrow{\spin{i}}}$ and $\overleftarrow{\spin{i}}$ is the set of spins whose indices are ``lexicographically'' less than $i$. $h_i(\spins)$ is a spatially \emph{local} or \emph{pointwise} measure and quantifies how much the individual spin \spin{i} contributes to the global entropy. We can more directly understand how local motifs contribute to the average entropy by plotting the local measures at each site within a given lattice configuration. This approach roughly parallels that in, \eg, Ref.~\cite{Lizi10a}, but here we employ different information measures and do not assume directionality, since we average the pointwise values resulting from each of the possible orientations centered at the given spin.

In the following, we define a \emph{cluster} to be a maximal set of contiguous spins oriented identically. The \emph{edge} of a cluster is that set of spins in a cluster with at least one neighbor not in the cluster. The remaining spins in the cluster are its \emph{bulk}. An isolated spin is a cluster consisting of a single spin.

\begin{figure}
  \centering
  \includegraphics{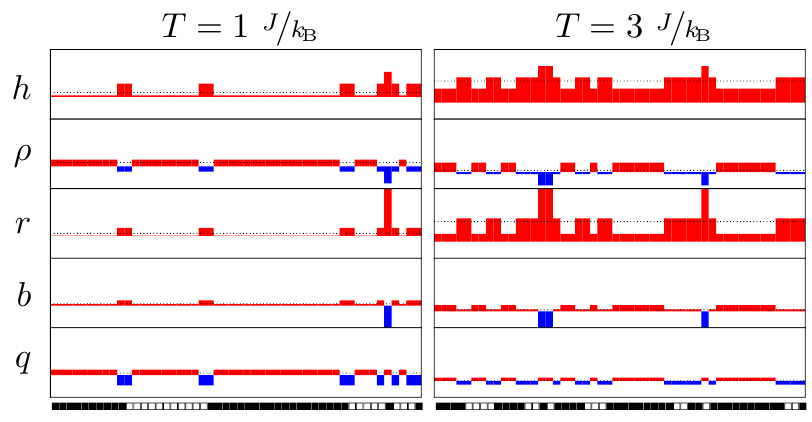}
  \caption{%
    Motif entropy-component analysis of the 1D Ising model at two temperatures. A segment of a spin configuration showing up spins (white cells) and down spins (black cells) is shown at the bottom.
  }
  \label{fig:ising_1d_local}
\end{figure}

\begin{figure}
  \centering
  \contourlength{0.125em}
  \includegraphics{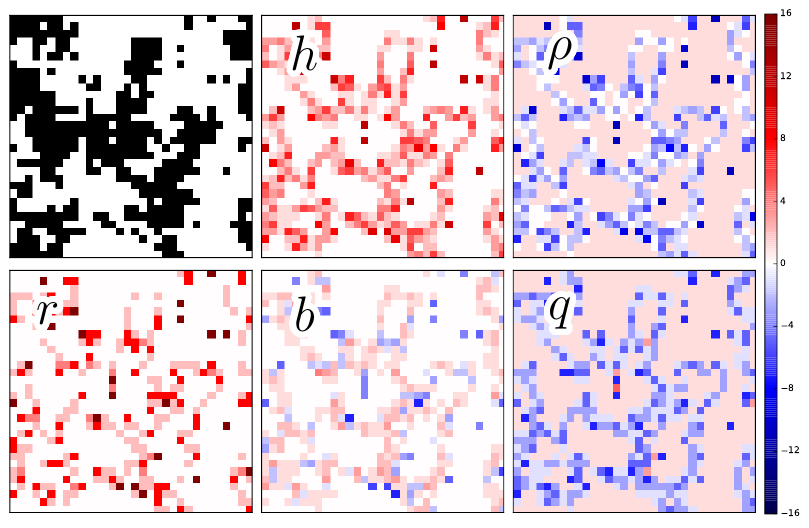}
  \caption{Motif entropy-component analysis of the 2D Ising model at \Tb.}
  \label{fig:ising_2d_local}
\end{figure}

Figures\nobreakspace \ref {fig:ising_1d_local} and\nobreakspace  \ref {fig:ising_2d_local} show the results of the motif entropy analysis in 1D and 2D lattices. In all cases, the spatial average of the displayed quantities corresponds (in the thermodynamic limit) with the values reported in Figs.\nobreakspace \ref {fig:ising_1d} and\nobreakspace  \ref {fig:ising_2d}.

There are several immediate similarities that allow for structural interpretation. For example, motif $q$ is positive within the bulk of a cluster and negative on its edges. Highlighting opposite features, motif $r$ is negligible within the bulk, but positive along edges, particularly corners and isolated spins. The motif \PATTERN entropy $b$, however, is more nuanced in its behavior. Considering Figure\nobreakspace \ref {fig:ising_1d_local}, at lower temperatures it is sensitive to cluster edges more so than cluster bulk. At higher temperatures, however, this relationship flips and it is more sensitive to the bulk. For all temperatures, it is negative for isolated spins---clusters of size $1$.

In contrast, we see that at not-too-high temperatures motif $b$ is largely sensitive to cluster boundaries. This leads one to speculate that, as in Ref.~\cite{barnett2013information}, $b$'s maximization in the disordered phase is due to complex interactions between cluster sizes and their surface area. This may also shed light on why $b$ does in fact peak at \Tc on the Bethe lattice. On the square lattice, on the one hand, cluster boundaries have a tendency to smooth due to the presence of correlations flowing along short loops in the lattice topology that tend to align spins on a boundary. On the other, the Bethe lattice has no such loops, and so there is no pressure to reduce surface area. This suggests that $b$ will generically peak at \Tc for systems where boundaries are not constrained by system energetics and that the energetic smoothing of cluster boundaries drives $b$ to peak in the more ``textured'' disordered phase.

Interestingly, for general lattices such as the Bethe lattice, but especially for random graph topologies, local (conditional) forms of the thermodynamic entropy density (analogs of Eq. (\ref{eq:localentropy})) and other motif measures are unknown and may simply not exist. While techniques, such as Ref.~\cite{jerrum1993polynomial}'s, exist for estimating thermodynamic entropy density for a ferromagnetic Ising model on an arbitrary graph, the interpretation of a thermodynamic entropy density in such systems is problematic as each spin site may have differing connectivity.  Therefore, while global averages may exist for arbitrary topologies and may even be tractably estimated, their structural meaning is vastly more challenging.

And, this is critical to understanding thermodynamic and statistical mechanical properties of such systems and, more generally, correlations and information measures on network dynamical systems. If local information estimates, such as those used for the motif entropy analyses above, existed, then studies can be undertaken that determine, say, which nodes in a lattice or network graph contribute most to collective behavior. Failing such measures, though, one must be wary of claims to answer such questions. These concerns are touched on by the informational analyses of frustration in spin glasses by Robinson \etal~\cite{Robi11a}, whose local entropies are temporally averaged, spatially local entropies in a heterogeneous system. They are unlike our motif entropies, though, which are both spatially and temporally local and, therefore, have meaning only in a homogeneous system. More study is required, however.

\subsection{Actively Storing Information versus Communicating It} \label{subsec:info_transfer}

Assuming Glauber dynamics for the nearest-neighbor 2D Ising system, Barnett \etal~\cite{barnett2013information} computed a \emph{global transfer entropy}~\cite{Schr00a}---the average information communicated between the entire lattice and a single spin. They found that it peaks within the disordered phase at a temperature $\Tgl \approx 2.354 \nicefrac{J}{\kB}$. Although there are serious concerns with how the transfer entropy conflates two-spin (dyadic) and multispin (polyadic) dependencies~\cite{PhysRevLett.116.238701}, those concerns are not relevant in these Ising systems due to their known, dyadic Hamiltonian.

\begin{figure}
  \includegraphics{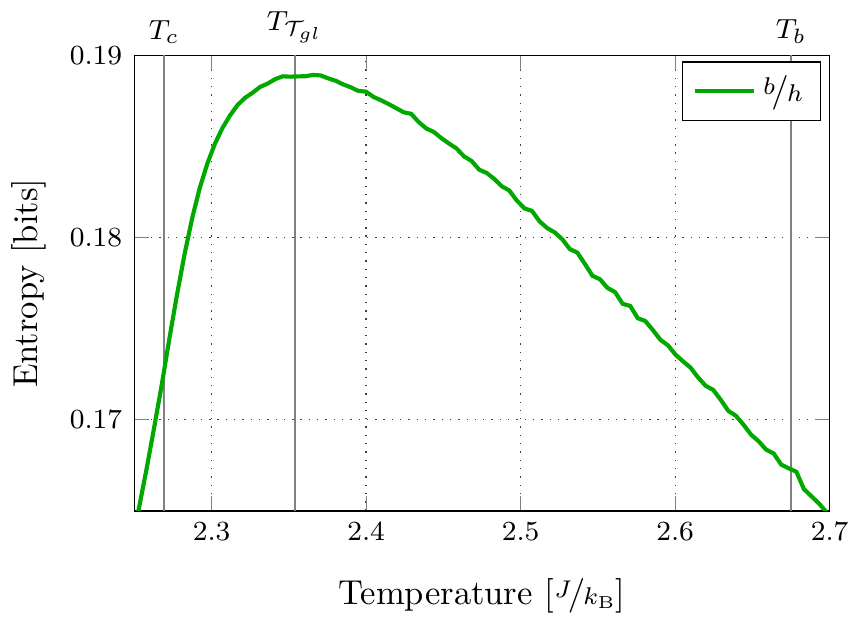}
  \caption{%
    Storing locally generated, thermal randomness as spatial correlation: The ratio of $\nicefrac{b}{h}$ as a function of temperature in the 2D Ising system. It peaks within the disordered phase at a temperature $\Tgl \approx 2.354 \nicefrac{J}{\kB}$. Compare to \cite[Fig.~2]{barnett2013information}.
	}
  \label{fig:b_over_h}
\end{figure}

To compare our measures with theirs, we examine the active storage of randomly generated information in spatial patterns, as measured by the ratio of $\nicefrac{b}{h}$---how much of the localized randomness ($h$) is converted by a system into spin-spin (spatially) shared information ($b$). Figure\nobreakspace \ref {fig:b_over_h} plots the ratio as a function of temperature for the 2D Ising system: the ratio peaks in the disordered phase and, surprisingly, it appears to do so at \Tgl. This implies that there is a strong connection between $b$ (storing thermal fluctuations as spatial correlation) and the potential for information communication in a system. Hopefully, the result suggests that Barnett \etal's \Tgl is not a result specific only to their choice of Glauber dynamics, but rather an intrinsic property of the 2D Ising model. It remains to be seen why spatially shared information is maximized within the disordered phase of the 2D Ising system, though.

\section{Conclusions}
\label{sec:conclusions}

As noted at the beginning, even the earliest debates over entropy's statistical foundation turned on contrasting its thermodynamic components---Boltzmann's isolated-spin entropy versus the Gibbs global entropy. From a modern perspective their difference, well known to Gibbs, is a generalized mutual information that measures the degree of individual-component independence, now called the total correlation. Keying off this, we showed that the Gibbs thermodynamic entropy density naturally decomposes further into two functionally distinct informational components: one quantifying independence among constituent spins and the other, dependence. The one quantifying dependence, the \PATTERN entropy $b$, captures collective behavior by expressing how much of the thermodynamic entropy density is locally shared. We then demonstrated the behavior of the \PATTERN entropy and related quantities for the nearest-neighbor ferromagnetic Ising model on a variety of lattices. We found that it tends to a maximum at intermediate temperatures, though not always at the magnetic phase transition. Our analyses support our earlier hypothesis~\cite{Ara14a} that $q$, as the dominant component of the persistent mutual information~\cite{ball2010quantifying}, should generically be maximized at critical points. Though not detailed here, this observation holds in three dimensions and in simulations of the Potts model~\cite{potts1952some} with the number of states $s = 3$ on lattices with $1$, $2$, and $3$ dimensions.

This brief phenomenological study of thermodynamic entropy components served to give a physical grounding for information measures and what they reveal in spin systems on various lattice topologies. The results suggest many avenues of potential research. One topic to explore is the behavior of $b$ in the Potts model, which switches from exhibiting a second-order phase transition in the magnetization to a first-order transition when the number of spin states exceeds four. Another setting of particular interest is to study the behavior of $b$ in a frustrated system, such as the antiferromagnetic Ising model on a triangular lattice, paralleling Ref.~\cite{Robi11a}. Finally, as alluded to above, although beyond the present scope of this work, a next step is to consider informational measures for spins on arbitrary graphs, with the goal of providing insight into the roles that different nodes play in information processing and storage in complex dynamical networks.

\section*{Acknowledgments}
\label{sec:acknowledgments}

We thank Cina Aghamohammadi, Adam Dioguardi, Raissa D'Souza, Pierre-Andr{\'e} No{\"e}l, M{\'a}rton P{\'o}sfai, Paul Riechers, Adam Rupe, Rajiv Singh, and Dowman Varn for helpful conversations. We thank the Santa Fe Institute for its hospitality during visits. JPC is an SFI External Faculty member. This material is based upon work supported by, or in part by, the U. S. Army Research Laboratory and the U. S. Army Research Office under contracts W911NF-13-1-0390 and W911NF-13-1-0340.

\appendix

\section{Bethe Lattice Spin-Neighborhood Probabilities}
\label{sec:bethe_analytic}

Consider a Cayley tree with coordination number $k$ consisting of $n$ shells, or layers of spins centered at the root. The total number of spins in such a tree is $N = k[ (k-1)^n - 1]/(k-2)$. We denote the spins in the lattice using \spin{i}, $i \in [0, N-1]$. Let us use the index $0$ for the central spin and indices $[1,k]$ for its immediate neighbors. We use \spins to denote a configuration---the state of all spins present in the Cayley tree. The Bethe lattice is the singular limit of a Cayley tree as $n \to \infty$. Following Ref.~\cite{baxter1982}, we write the partition function for the Ising model on the Bethe lattice as:
\begin{align*}
  Z = \sum_{\mathclap{\spins \in \spinsrv}} \exp(\beta J \sum_{\langle i,j \rangle} \spin{i} \spin{j})
  ~,
\end{align*}
where $\exp(\beta J \sum_{\langle i,j \rangle} \spin{i} \spin{j})$ is the Boltzmann factor corresponding to the system being in a state \spins and \spinsrv denotes the set of all possible configurations. As shown in Ref.~\cite{baxter1982}, due to the Cayley tree topology, the above expression can be written as:
\begin{align*}
  Z = \sum_{\mathclap{\spin{0} \in \spinrv{0}}} g_n(\spin{0})^k
  ~,
\end{align*}
where $g_n(\spin{0})$ is the partition function of a branch of the Cayley tree with its root at $\spinrv{0}$. Starting with this expression we derive the joint probability of the central spin and its neighbors.

Let $g_n(\spin{0}) \to g(\spin{0})$ in the limit of $n \to \infty$ shells. Then, the spins not \emph{close} to the Cayley tree leaves behave like those on the Bethe lattice with the same coordination number. In this limit, by explicitly accounting for the bonds between the central spin and its neighbors, we can write the partition function as:
\begin{align*}
  Z = \sum_{{\substack{\spin{0} \in \spinrv{0} \\
                                \spin{1} \in \spinrv{1} \\
                                \vdots \\
                                \spin{k} \in \spinrv{k}}}}
    \left[ \exp \left( \beta J \sum_{\mathclap{i \in [1,k]}} \spin{0} \spin{i} \right)
      \prod_{{i \in [1,k]}} [g(\spin{i})]^{k-1} \right]
  ~,
\end{align*}
where the product is essentially over all branches 1 step away from the central spin, and \spin{i} is the spin to which it is anchored. From this, we obtain the joint probability of the central spin \spin{0} and its $k$ neighbors as:
\begin{align*}
  p(\spin{0}, & \spin{1}, \ldots, \spin{k}) \\
  & =  \exp \left( \beta J \sum_{\mathclap{i \in [1,k]}} \spin{0} \spin{i} \right)
  	\prod_{{i \in [1,k]}} [g(\spin{i})]^{k-1} / Z
  ~.
\end{align*}
Dividing both numerator and denominator by $[g(+1)]^{k-1}$ we obtain:
\begin{widetext}
  \begin{align}
    p(\spin{0}, \spin{1}, \ldots, \spin{k}) =  \frac{\displaystyle \exp \left( \beta J \sum_{\mathclap{i \in [1,k]}} \spin{0} \spin{i} \right) \prod_{{i \in [1,k]}} \left[ \frac{g(\spin{i})}{g(+1)} \right]^{k-1}}
    {\displaystyle \left[ e^{\beta J} + e^{- \beta J} \left[ \frac{g(-1)}{g(+1)} \right]^{k-1} \right]^k + \left[ e^{- \beta J} + e^{ \beta J} \left[ \frac{g(-1)}{g(+1)} \right]^{k-1} \right]^k }
    ~.
  \end{align}
\end{widetext}
This is the joint probability distribution of the central spin and its neighbors. We evaluate the above expression numerically by defining $g(-1)/g(+1) = x$. From Ref. \cite[Eq.~(4.3.14)]{baxter1982}, we know that $x$ is the ``stable'' root of the equation:
\begin{align}
  x = \frac{e^{- \beta J } + e^{\beta J } x^{k-1} }
  { e^{ \beta J } + e^{ - \beta J } x^{k-1}}
  ~.
  \label{eq:x}
\end{align}

Above \Tc, Eq.\nobreakspace \textup {(\ref {eq:x})} has only one root: $x=1$, which is stable. Below \Tc, however, there are there are 3 roots: $x_0, 1, {x_0}^{-1}$ where $x_0 < 1$. The stable roots are $x_0$ and ${x_0}^{-1}$. Setting $x = x_0$ provides the distribution $p(\spin{0}, \spin{1}, \ldots, \spin{k})$ where the symmetry breaking prefers spins aligned up and where setting $x = {x_0}^{-1}$ provides the distribution with the symmetry broken such that downward spins are preferred.


\begin{thebibliography}{10}

\bibitem{antonov2011electronic}
V.~N. Antonov, L.~V. Bekenov, and A.~N. Yaresko.
\newblock Electronic structure of strongly correlated systems.
\newblock {\em Adv. Cond. Matter Physics}, 2011(298928):1--207, 2011.

\bibitem{mantegna1999introduction}
R.~N. Mantegna and H.~E. Stanley.
\newblock {\em Introduction to Econophysics: {Correlations} and complexity in
  finance}.
\newblock Cambridge University Press, Cambridge, United Kingdom, 1999.

\bibitem{raiesdana2008complexity}
S.~Raiesdana, M.~R. Hashemi~Golpayegani, and A.~M. Nasrabadi.
\newblock Complexity evolution in epileptic seizure.
\newblock In {\em Engineering in Medicine and Biology Society, 2008. EMBS 2008.
  30th Annual International Conference of the IEEE}, pages 4110--4113. IEEE,
  2008.

\bibitem{maki2013disruption}
V.~M{\"a}ki-Marttunen, J.~M. Cortes, M.~F. Villarreal, and D.~R. Chialvo.
\newblock Disruption of transfer entropy and inter-hemispheric brain functional
  connectivity in patients with disorder of consciousness.
\newblock {\em BMC Neuroscience}, 14(Suppl 1):P83, 2013.

\bibitem{couzin2007collective}
I.~Couzin.
\newblock Collective minds.
\newblock {\em Nature}, 445(7129):715--715, 2007.

\bibitem{barnett2013information}
L.~Barnett, J.~T. Lizier, M.~Harr{\'e}, A.~K. Seth, and T.~Bossomaier.
\newblock Information flow in a kinetic {Ising} model peaks in the disordered
  phase.
\newblock {\em Phys. Rev. Lett.}, 111(17):177203, 2013.

\bibitem{Cove06a}
T.~M. Cover and J.~A. Thomas.
\newblock {\em Elements of Information Theory}.
\newblock Wiley-Interscience, New York, second edition, 2006.

\bibitem{Crut12a}
J.~P. Crutchfield.
\newblock Between order and chaos.
\newblock {\em Nature Physics}, 8(January):17--24, 2012.

\bibitem{Shaw84}
R.~Shaw.
\newblock {\em The Dripping Faucet as a Model Chaotic System}.
\newblock Aerial Press, Santa Cruz, California, 1984.

\bibitem{Arno96}
D.~Arnold.
\newblock Information-theoretic analysis of phase transitions.
\newblock {\em Complex Systems}, 10:143--155, 1996.

\bibitem{crutchfield1997statistical}
J.~P. Crutchfield and D.~P. Feldman.
\newblock Statistical complexity of simple one-dimensional spin systems.
\newblock {\em Phys. Rev. E}, 55(2):R1239, 1997.

\bibitem{feldman2003structural}
D.~P. Feldman and J.~P. Crutchfield.
\newblock Structural information in two-dimensional patterns: {Entropy}
  convergence and excess entropy.
\newblock {\em Phys. Rev. E}, 67(5):051104, 2003.

\bibitem{Feld08a}
D.~P. Feldman, C.~S. McTague, and J.~P. Crutchfield.
\newblock The organization of intrinsic computation: {Complexity}-entropy
  diagrams and the diversity of natural information processing.
\newblock {\em CHAOS}, 18(4):043106, 2008.

\bibitem{lau2013information}
H.~W. Lau and P.~Grassberger.
\newblock Information theoretic aspects of the two-dimensional {Ising} model.
\newblock {\em Phys. Rev. E}, 87(2):022128, 2013.

\bibitem{Jame11a}
R.~G. James, C.~J. Ellison, and J.~P. Crutchfield.
\newblock Anatomy of a bit: {Information} in a time series observation.
\newblock {\em CHAOS}, 21(3):037109, 2011.

\bibitem{Schr00a}
T.~Schreiber.
\newblock Measuring information transfer.
\newblock {\em Phys. Rev. Lett.}, 85:461--464, 2000.

\bibitem{abdallah2012measure}
S.~A. Abdallah and M.~D. Plumbley.
\newblock A measure of statistical complexity based on predictive information
  with application to finite spin systems.
\newblock {\em Phys. Lett. A}, 376(4):275--281, 2012.

\bibitem{verdu2006erasure}
S.~Verd{\'u} and T.~Weissman.
\newblock Erasure entropy.
\newblock In {\em Information Theory, 2006 IEEE Intl. Symp.}, pages 98--102.
  IEEE, 2006.

\bibitem{james2014chaos}
R.~G. James, K.~Burke, and J.~P. Crutchfield.
\newblock Chaos forgets and remembers: {Measuring} information creation,
  destruction, and storage.
\newblock {\em Phys. Lett. A}, 378(30):2124--2127, 2014.

\bibitem{jaynes1965gibbs}
E.~T. Jaynes.
\newblock Gibbs versus {Boltzmann} entropies.
\newblock {\em Am. J. Physics}, 33(5):391--398, 1965.

\bibitem{Shan48a}
C.~E. Shannon.
\newblock A mathematical theory of communication.
\newblock {\em Bell Sys. Tech. J.}, 27:379--423, 623--656, 1948.

\bibitem{grandy2008entropy}
W.~T. Grandy~Jr.
\newblock {\em Entropy and the time evolution of macroscopic systems}, volume
  141.
\newblock Oxford University Press, Oxford, United Kingdom, 2008.

\bibitem{ruelle1969rigorous}
D.~Ruelle.
\newblock {\em Statistical Mechanics: Rigorous results}.
\newblock World Scientific, Singapore, 1969.

\bibitem{Grim99a}
G.~R. Grimett.
\newblock {\em Percolation}.
\newblock Springer, Berlin, second edition, 1999.

\bibitem{reza1961introduction}
F.~M. Reza.
\newblock {\em An introduction to information theory}.
\newblock Courier Corporation, 1961.

\bibitem{goldstein1990entropy}
S.~Goldstein, R.~Kuik, and A.~G. Schlijper.
\newblock Entropy and global {Markov} properties.
\newblock {\em Comm. Math. Physics}, 126(3):469--482, 1990.

\bibitem{schlijper1989two}
A.~G. Schlijper and B.~Smit.
\newblock Two-sided bounds on the free energy from local states in {Monte
  Carlo} simulations.
\newblock {\em J. Stat. Phys.}, 56(3-4):247--260, 1989.

\bibitem{schurmann2015note}
T.~Sch{\"u}rmann.
\newblock A note on entropy estimation.
\newblock {\em arXiv preprint arXiv:1503.05911}, 2015.

\bibitem{wolff1989collective}
U.~Wolff.
\newblock Collective {Monte Carlo} updating for spin systems.
\newblock {\em Phys. Rev Lett.}, 62(4):361, 1989.

\bibitem{maniwa2003}
M.~Yutaka, H.~Kataura, K.~Matsuda, and Y.~Okabe.
\newblock A one-dimensional {Ising} model for {C 70} molecular ordering in {C
  70}-peapods.
\newblock {\em New J. Physics}, 5(1):127, 2003.

\bibitem{zimmermann2000}
F.~M. Zimmermann and X.~Pan.
\newblock Interaction of ${H}_{2}$ with
  $\mathrm{Si}(001)-(2\ifmmode\times\else\texttimes\fi{}1)$: Solution of the
  barrier puzzle.
\newblock {\em Phys. Rev. Lett.}, 85:618--621, 2000.

\bibitem{bruno1960}
B.~H. Zimm.
\newblock Theory of ``melting'' of the helical form in double chains of the
  {DNA} type.
\newblock {\em J. Chem. Physics}, 33(5):1349--1356, 1960.

\bibitem{wartell1985}
R.~M. Wartell and A.~S. Benight.
\newblock Thermal denaturation of {DNA} molecules: {A} comparison of theory
  with experiment.
\newblock {\em Phys. Reports}, 126(2):67 -- 107, 1985.

\bibitem{Durl99a}
S.~N. Durlauf.
\newblock How can statistical mechanics contribute to social science?
\newblock {\em Proc. Natl. Acad. Sci. USA}, 96(19):10582--10584, 1999.

\bibitem{pathria1996}
R.~K. Pathria and P.~D. Beale.
\newblock {\em Statistical Mechanics}.
\newblock Elsevier Science, Amsterdam, Netherlands, 1996.

\bibitem{baxter1982}
R.~J. Baxter.
\newblock {\em Exactly solved models in statistical mechanics}.
\newblock Academic Press, New York, New York, 1982.

\bibitem{pfeuty1979}
P.~Pfeuty.
\newblock An exact result for the {1D} random {Ising} model in a transverse
  field.
\newblock {\em Phys. Lett. A}, 72(3):245--246, 1979.

\bibitem{Feld98a}
D.~P. Feldman.
\newblock {\em Computational Mechanics of Classical Spin Systems}.
\newblock PhD thesis, University of California, Davis, 1998.
\newblock {P}ublished by University Microfilms Intl, Ann Arbor, Michigan.

\bibitem{yilmaz2005}
M.~B. Yilmaz and F.~M. Zimmermann.
\newblock Exact cluster size distribution in the one-dimensional {Ising} model.
\newblock {\em Phys. Rev. E}, 71:026127, Feb 2005.

\bibitem{bollobas2001}
B{\'e}la Bollob{\'a}s.
\newblock {\em Random Graphs, volume 73 of Cambridge studies in advanced
  mathematics}.
\newblock Cambridge University Press, Cambridge,, 2001.

\bibitem{marcus2013computing}
B.~Marcus and R.~Pavlov.
\newblock Computing bounds for entropy of stationary {Z}$^d$ {Markov} random
  fields.
\newblock {\em SIAM J. Discr. Math.}, 27(3):1544--1558, 2013.

\bibitem{Lizi10a}
J.~Lizier, M.~Prokopenko, and A.~Zomaya.
\newblock Information modification and particle collisions in distributed
  computation.
\newblock {\em CHAOS}, 20(3):037109, 2010.

\bibitem{jerrum1993polynomial}
M.~Jerrum and A.~Sinclair.
\newblock Polynomial-time approximation algorithms for the {Ising} model.
\newblock {\em SIAM J. Computing}, 22(5):1087--1116, 1993.

\bibitem{Robi11a}
M.~D. Robinson, D.~P. Feldman, and S.~R. McKay.
\newblock Local entropy and structure in a two-dimensional frustrated system.
\newblock {\em Chaos}, 21:037114, 2011.

\bibitem{PhysRevLett.116.238701}
R.~G. James, N.~Barnett, and J.~P. Crutchfield.
\newblock Information flows? a critique of transfer entropies.
\newblock {\em Phys. Rev. Lett.}, 116:238701, Jun 2016.

\bibitem{Ara14a}
P.~M. Ara, R.~G. James, and J.~P. Crutchfield.
\newblock The elusive present: {Hidden} past and future dependence and why we
  build models.
\newblock {\em Phys. Rev. E}, 93(2):022143, 2016.

\bibitem{ball2010quantifying}
R.~C. Ball, M.~Diakonova, and R.~S. MacKay.
\newblock Quantifying emergence in terms of persistent mutual information.
\newblock {\em Adv. Complex Systems}, 13(03):327--338, 2010.

\bibitem{potts1952some}
R.~B. Potts.
\newblock Some generalized order-disorder transformations.
\newblock In {\em Math. Proc. Cambridge Phil. Soc.}, volume~48, pages 106--109,
  Cambridge, United Kingdom, 1952. Cambridge University Press.

\end{thebibliography}
\end{document}